\documentstyle[aps,prl]{revtex}
\draft
\begin{document}
\title {Luminous matter may arise from
a turbulent plasma state of the early universe.}
\author {Per Bak{$\ddagger$} and Maya Paczuski{$\dagger$}}
\address{{$\ddagger$} 
Niels Bohr Institute, Blegdamsvej 17, Copenhagen, Denmark.\\
{$\dagger$}
Department of Mathematics, Imperial College, London SW7 2BZ, UK.}
\date{\today}
\maketitle

{\bf The almost perfect uniformity of the cosmic microwave background (CMB) radiation,
discovered by Penzias and Wilson in 1965 \cite{penzias}, appears to present clearcut evidence that the universe was uniform and in equilibrium at the decoupling transition when a plasma of protons and electrons condensed into a gas of Hydrogen.  COBE indicates that only very small ripples of order $10^{-5}$ existed at decoupling.
Gravity then caused hydrogen to cluster  and possibly reheat parts of the universe to form the luminous matter that we observe today.

We suggest an alternative scenario, where a spatially intermittent structure of extremely hot matter already existed in an otherwise
uniform plasma state at the decoupling transition. The  plasma was not in equilibrium but in a very high Reynolds number turbulent state. The
sparse bursts would not affect the uniformity of the CMB radiation.
Luminous matter originates from
localized hot bursts already present in the plasma state
prior to decoupling. No reheating, and no exotic matter is needed to get luminous matter.}

The decoupling transition occurred when the temperature of the universe
reached Hydrogen's ionization energy, $T_c=4,200 K$ degrees approximately 300,000 years after the Big Bang.  Since radiation interacts with the electrons in the plasma, but not with the Hydrogen, the  radiation thereafter simply cooled uniformly, reducing the temperature to the much smaller value $T=2.736 K$ that we observe today. Since the radiation
was in equilibrium with the plasma at $T_c$, the uniformity of the CMB radiation
implies that
the gas of Hydrogen was also almost perfectly uniform at that time.

In contrast, the observed structure of luminous matter is strongly clustered, intermittent, and fractal-like, with correlations over
perhaps hundreds of millions of light years \cite{labini,bak}.
Although some clustering is observed, computer simulations have been unable to explain the large scale structure of luminous matter using well-established physics, starting from a uniform equilibrium state with small perturbations or ripples. For instance, the
Hubble Volume Project includes hypothetical Cold Dark Matter in order to obtain reasonable agreement\cite{hubble1}\cite{hubble2}\cite{hubble3}.

Thinking about the problem inversely, there has not been time enough
for the large scale geometry of luminous or hot matter at $T_c$ to differ very
much from today's. Thus, luminous or hot matter had to be present already
at $T_c$, with an intermittent geometry. In fact, intensive luminous matter existed at least as early as fourteen billion years ago\cite{lanzetta}, only one billion years after $T_c$.
How can that be reconciled with the observed uniformity of the
CMB radiation?

It is enough to assume that the plasma before $T_c$ was in a turbulent
state, with a very high Reynolds number.
 Actually, turbulent plasmas are plentiful in the universe,
for instance in the sun.  
The properties of turbulent systems are determined
by the dimensionless Reynolds number, $R=Vl_0/\nu$, where $\nu$ is the
viscosity, $V$ the velocity difference over the integral scale $l_0$, which may be large, for instance the entire system.

For very large $R$, velocity gradients are extremely small almost everywhere. Regions with vanishingly small velocity gradients
are in effective thermal equilibrium, since there is almost no dissipation or heating there. However, the energy injected at the scale $l_0$ is dissipated over extremely small scales, the Kolmogorov length $l_0 R^{-3/4}$, or even smaller length scales within the so-called intermediate dissipative range. The velocity gradients in these localized, sparse, filamentary regions are enormous.  In the
bursts, the plasma is heated to temperatures that are very much higher than the  temperature in the surrounding 
"equilibrium" plasma sea. The dissipative field is highly intermittent, with "on" regions 
having a fractal or fractal-like structure \cite{frisch}, embedded in
the "off" background of no dissipation.
 Note that for smaller Reynolds number, there is no clear distinction between dissipating and
nondissipating regions.  The distinction becomes sharper and sharper as the Reynolds
number increases. In particular  the characteristic length
scale of the regions with large dissipation  (the bursts) decreases, and the amplitude of the dissipation in the background  outside of the bursts decreases as well, approaching the equilibrium limit of no dissipation.  Thus turbulence becomes an "on/off" phenomena
when the Reynolds number is very large.

 For an example of intermittent structures with high Reynolds number, of order $10^5$, that can be achieved on earth, see the work by Meneveau and Sreenivasan on the atmospheric boundary layer\cite{srini}. Or think of flying across the
Atlantic on a clear calm day: suddenly and unpredictably the plane may
enter a very violent burst. The burst ends as suddenly as it starts.

This might well describe the plasma before $T_c$. The Reynolds number would  be huge even for a small value of $V$, because of the large value of $l_0$, being of order the size of the universe.It is, however
very difficult to make numerical estimates; even on the more directly observable earthly
conditions intermittent turbulence is difficult to calculate and descibe.

What happens as the plasma cools to $T_c$? Over the vast, almost space filling, equilibrium volume, the plasma suddenly condenses into a Hydrogen gas, decoupling mass and radiation, thereby creating the uniform CMB radiation. But
the hot, sparse bursts pass through the transition without
change, since their much higher temperatures prevent
condensation. The CMB radiation is unaffected due to
the sparsity of these hot bursts embedded in an equilibrium sea.  For an analogous situation, consider the recent universe where the microwave field
is certainly unaffected by the existence of the intermittent luminous
matter. Gibson\cite{gibson} has suggested that turbulence existed
in the very early universe, but there is no hot plasma at $T_c$ in his picture.

We suggest these hot intermittent bursts in the plasma at $T_c$ were responsible for the
frenzied star formation soon after $T_c$\cite{lanzetta}, and evolved into the stars and galaxies that we see today, while the microwave field cooled to $T=2.736K$. The small fluctuations or ripples observed by COBE are insignificant for this scenario. The rarefied, intermittent bursts of hot matter
in the equilibrium sea cannot affect the CMB radiation. Also, there is no reheating or exotic physics necessary in order to explain the existence of luminous matter.

Luminous matter may have evolved from rare, hot
bursts already existing in the turbulent plasma of the early universe. It may be no accident
that the current structure of luminous matter has close similarities with that of very high Reynolds number turbulence.

\end{document}